\documentstyle[12pt]{article}
\begin{document}
\centerline{{\huge Electronic Shell Structure}}
\medskip
\centerline{{\huge of Large Metallic Clusters}}
\medskip
\centerline{{\huge in the Modified Harmonic Oscillator }}
\medskip
\centerline{W.D. Heiss$^{\star}$\footnote{heiss@physnet.phys.wits.ac.za}
and R.G. Nazmitdinov$^{\star \star}$}
\medskip
\centerline{
$^{\star}$ Centre for Nonlinear Studies and Department of Physics}
\centerline{University of the Witwatersrand, PO Wits 2050,
Johannesburg, South Africa}
\centerline{$^{\star \star}$ Bogoliubov Laboratory of Theoretical Physics}
\centerline{Joint Institute for Nuclear Research, 141980 Dubna, Russia}
\baselineskip 20pt minus.1pt
\begin{abstract}
Shell structure in the single particle
spectrum of deformed harmonic oscillator
potentials when a term proportional to $(\vec L)^2$ is added is analyzed
for a large particle number. In the case
study presented here it is argued that, in view of the chaotic nature of the
problem, a thorough understanding of the classical situation provides
essential guidance in tackling the corresponding quantum mechanical
problem. A scaling law which gives a dividing line between regular and
chaotic behavior in terms of energy, deformation and strength of the
$(\vec L)^2$ term has been found. According to this law, shell structure
survives for higher particle numbers only with lesser deformation.
\end{abstract}
\vspace{0.2in}
\vspace{0.2in}

\section{Introduction}
Finite Fermi systems give rise to shell structure as is well known from
the existence of magic numbers in atoms and nuclei.
Nowadays, shell effects are also established in
metallic clusters \cite{He93,Bra93,Com}.
The first experimental evidence in alkali metallic
clusters \cite{Kn}, and independently the theoretical prediction of magic
numbers within the framework of the spherical jellium model
\cite{Ek,Be}, have been major hallmarks in this field.

The valence electrons in metallic clusters are strongly
delocalized. They play an essential role in the understanding
of many observables and they can be treated by the mean field approach.
The jellium model works very effectively for monovalent metallic clusters
like sodium and potassium. This mean field model describes a single valence
electron that interacts with the average potential which is
generated by the other electrons and ions.
The structure of the ions is neglected, they are assumed to form
a constant positive background or jellium density
within a finite volume. While the spherical jellium
model has been able to explain general features of enhanced abundances
of valence electrons in spherical clusters \cite{Bj90}--\cite{Ma91},
it has been recognized that the finer structure in the
mass spectra between magic numbers can be explained via symmetry breaking
mechanisms \cite{Cl85}--\cite{RBH} similar to the situation in nuclear
physics \cite{BM75}. A spherically symmetric mean field leads to very
strong shell effects which is manifested in the stability of the noble gases,
magic nuclei and magic metallic clusters. When a spherical shell is
only partially filled, a spontaneous breaking of spherical symmetry
can give rise to a deformed equilibrium shape, if it is associated with
an energy gain. For clusters with incompletely filled shell, it has been
suggested, in analogy with atomic nuclei, to consider a quadrupole
deformation which leads to the lifting of the degeneracies in the spherical
model. In fact, this idea was implemented by Clemenger \cite{Cl85} who
applied the modified harmonic oscillator model (MHO) to metallic clusters
where the model played an important role in the understanding of various
experimental data (see for review \cite{He93,Bra93}). The MHO with
a spin--orbit term was introduced
by Nilsson \cite{N} in nuclear physics, and it is a
very effective tool in the analysis of ground state deformations
and single particle excitations for many nuclei \cite{BM75,NR}.
The same model without the spin--orbit term
seems to work successfully in reproducing the shapes of
small clusters which have been calculated
by the more sophisticated methods of quantum chemistry \cite{He93}.
Based on these results it is therefore accepted that cluster
deformations exist. This is in fact confirmed at least for
clusters with $A\leq 40$ either by the Clemenger--Nilsson (CN) model
or by a self-consistent
Kohn--Sham density--functional method \cite{KS65} (KS) with deformed jellium
backgrounds \cite{EP}--\cite{Mon95}.

Nowadays there is experimental evidence of shells for metallic
clusters with more than 1000 atoms \cite{Bj90}--\cite{Ma91},\cite{Ped,Br}.
The shell ground state properties
of spherical metallic clusters containing a few
thousand valence electrons can be described within the KS approach using
the spherical jellium model \cite{Gen}. Solving the KS equations is quite
difficult when deformation effects are taken into account, even for modern
supercomputers. It turns out that phenomenological potentials
used traditionally in nuclear physics serve a purpose similar to those
obtained within the KS approach if the relevant parameters are adjusted
appropriately. Typical potentials are the
Woods--Saxon (and its various modifications) (WS) and
the MHO potential, i.e.
the CN model. The main idea of the approach exploiting the WS and the MHO is
to use the single particle spectrum generated by the potential to calculate
shell energies only, because neither WS nor CN models allow for a
description of binding energies for neutral or charged clusters.
Following Strutinsky's idea \cite{Str}, the energy of an
interacting Fermi system is divided into a smooth part, denoted
by $\tilde{E}$, and an oscillating part which is
the shell correction energy denoted by $\delta E$, i.e.
$E=\tilde{E} + \delta E$. The quantity $\tilde E$ varies slowly with particle
number and the deformation of the system. In nuclear physics and for metallic
clusters it is usually replaced by the liquid drop energy
which carries the information about the bulk properties of the system.
The shell correction energy $\delta E$ contains all the oscillations
originating from the bunching of the energy levels. It is obtained by
summing the single particle energies of a phenomenological shell model
potential and subtracting the average smooth part of the total
energy \cite{Str,BM75,NR}.

Using the spherical WS potential for clusters with
thousand valence electrons Nishioka {\it et al} \cite{Ni90}
predicted the super-shell phenomenon where the regular oscillations
in the level density which reflect the main shells are modulated
by an amplitude which oscillates with a lower frequency. Such analysis led
to the experimental discovery of the most stable large spherical metallic
clusters \cite{Ped,Br,Com}. Employing the Strutinsky shell correction method,
a systematic study of cluster equilibrium shapes has
been performed for the WS potential \cite{Fr93,FP} up to 300 atoms and for
the MHO potential for clusters with particle numbers up
to 300 in \cite{Ya95,RFB} and up to 800 in \cite{RBH}.
Effects of deformation are clearly seen in these calculations.
For clusters with particle number of up to a few thousand
it is natural to use these potentials
for the analysis of the existence of deformation.
The problem is still being discussed in the ongoing
literature (see, for example, discussions by S.Bj\o rnholm in \cite{Com} and
\cite{Bul}).

The equilibrium deformation is ultimately related to the behavior
of the shell correction energy $\delta E$
of the quantum Hamiltonian. According to the periodic orbit theory
\cite{Gut}-\cite{S76} (see also \cite{BM75,BrB})
the frequencies in the oscillations of quantum single particle
spectra are determined by the corresponding periods of classical
closed periodic orbits. The gross shell structure is due to
the periodic orbits with a shortest period, whereas the finer details are
determined by contributions of longer orbits.
This result has been confirmed for spherical potentials
used for large clusters \cite{Ni90,Ma93,Le93,Pa93}.
It suggests that the semi-classical approach can provide deeper insight
also into the problem of large deformed clusters
by exploiting the tools which have been developed in the study
of classical integrable and non-integrable systems.

One of the first attempts to analyze the interplay between shell effects and
classical chaotic motion has been done in \cite{Ar87} for an
ellipsoidally deformed version of a diffuse nuclear average potential called
the Buck-Pilt potential. If the classical problem is
non-integrable and displays chaotic behavior, the shell structure of
the corresponding quantum spectrum is affected depending on the degree of
chaos. More recently these effects were demonstrated for the
quadrupole + octupole \cite{H94},
quadrupole + hexadecapole \cite{H95} and for
quadrupole + octupole + hexadecapole \cite{HNR} axial deformed harmonic
oscillator potentials.

Initiated by these latter investigations we make use of the
connection between ordered motion in the classical problem and shell
structure phenomena in the corresponding quantum spectra for
phenomenological potentials used for clusters. In practical applications
the study of the more realistic deformed WS potential is complicated due to
the effective occurrence of many multipoles; this is the subject of a future
paper. Since the MHO is being used widely for the interpretation of data
\cite{He93,RBH,Ya95,RFB,Eat} we expand in the present paper the preliminary
classical analysis of the MHO potential \cite{hena} and
demonstrate its connection with quantum--mechanical
single particle orbitals. One particular result of the
classical analysis is a kind of scaling relationship between energy range and
the strength $\lambda$ of the $(\vec L)^2$ term in the MHO potential.
For the corresponding quantum mechanical problem this translates into the
finding that pronounced shell structure can occur for increasing
particle numbers only for an accordingly decreasing strength $\lambda$
and/or a decreasing quadrupole deformation.

\section{General properties of the model}

The level ordering in the WS potential
falls between the soft-surface harmonic oscillator (HO) and the hard-surface
square well. A similar level ordering is obtained by modifying the HO.
The MHO reads in cylindrical coordinates
\begin{equation}
\label{nil}
H={1\over 2m}(p_{\varrho }^2+p_z^2+{p_{\varphi }^2\over \varrho ^2})
+{m \omega ^2\over 2}(\varrho ^2+{z^2\over b^2}) -
\lambda \hbar \omega (\vec L)^2
\label{ham} \end{equation}
where $\vec L$ is the angular momentum in units of $\hbar$ and
$\lambda >0$.
 As the model was originally designed within a pure quantum
mechanical context \cite{BM75,N} we conveniently use the energy unit
$\hbar \omega $ for the $(\vec L)^2$-term. We consider
only axial quadrupole deformations. For $b=1$ all three components
of $\vec L$ are conserved because of spherical symmetry; the model
becomes then integrable. For $b>1$ and $b<1$, which is referred
to as prolate and oblate quadrupole deformation, respectively, the problem is
non-integrable. The axial symmetry makes the problem a two degrees of
freedom system with $L_z$, denoted classically by $p_{\varphi }$, being
a constant of motion.

The classical phase space is unbounded for given energy.
The subtraction of a term proportional to the
square of the angular momentum allows in principle arbitrarily large
values of positions and momenta. This is easily recognized for $b=1$ where
an arbitrarily large value of the then conserved quantity
$(\vec L)^2$ can be given as one of the initial condition;
using correspondingly large values of positions and/or momenta, an
arbitrary value of the energy can be attained.
The corresponding quantum mechanical levels are given
by the expression $E_N \approx \hbar \omega l(1-\lambda l)$
for $n_r=0\, (N=2n_r+l)$ for sufficiently large $l$.
For $l>1/\lambda$ the energy levels are therefore negative. This
unboundedness has been dealt with tentatively by adding to the Hamiltonian
a term proportional to $\langle N|L^2|N\rangle $. The expectation value is
taken with shell model states and can be expressed in terms of
shell numbers \cite{BM75,NR}, if such quantum numbers are well defined
which applies to the regular region only. For $b\ne 1$,
when $(\vec L)^2$ is no longer conserved, its time variations extent over a
fairly large range of values for generic orbits;
depending on initial conditions such values have no bound in principle.
We enter the chaotic region and the corresponding quantum mechanical
states become mixtures of angular momentum states. As long as the range
of such admixtures obeys the inequality $l\leq l_{{\rm crit}}$ with
$\lambda\cdot l_{{\rm crit}}=1$ for a given value of $\lambda$, the
states are not affected by the unboundedness of the Hamiltonian (\ref{nil}).
It is discussed below that, for $b>1$ and the range of $\lambda $-values
considered, such states give rise to the chaotic nature of the spectrum.
In contrast, states with large values of $l$ exceeding $l_{{\rm crit}}$
are basically regular and do not mix with the former group of states. The
latter states give rise to negative energies but are irrelevant for our
discussion as they are essentially orthogonal to the former relevant group.

Since the model is non-integrable for $b\ne 1$ and $\lambda \ne 0$
there is no obvious basis for the corresponding
quantum mechanical problem. If $\lambda \ll 1$ and $b>1$ the
adequate basis makes use of the asymptotic quantum numbers \cite{NM} which
diagonalize the Hamiltonian for $\lambda =0$ (see below).
While this approach facilitates the technical problem,
even this representation requires great numerical effort
to tackle the problem. In accordance with the classical problem,
the quantum spectrum shows all characteristics of chaotic
behavior for large systems.

We have chosen the upper limit $b=2$, which is denoted
super-deformation for nuclei \cite{NR}. Super-deformation seems
to exist in metallic clusters, at least for small particle
numbers (see, for example \cite{He93,Bra93,Hir94,Mon95,FP,Ya95,RFB}).
We focus our attention only on $1< b\leq 2$, since $b<1$ (oblate case)
can be obtained from the results of the former case by interchanging
$\varrho $ and $z$ and suitably replacing $\omega $ by $ b\omega $.
This is in stark contrast with the situation considered in
\cite{Ar87,H94,HNR}. There the oblate case is chaotic and shell structure
is essentially destroyed once higher multipoles are added while shell
structure can prevail in the prolate case. Note, that the potential used in
\cite{Ar87} contains effectively higher multipoles.

\section{Classical treatment}

For most of the results presented here
it suffices to consider $p_{\varphi }=0$ in Eq.(\ref{ham}). Larger values
of $p_{\varphi }$ are discussed at the end of this section.

We first consider qualitatively the case of
spherical symmetry, i.e.$\,b=1$.
The problem reduces to one degree of freedom since now
$p_{\theta }=m\dot \theta r^2=z p_{\varrho }-\varrho p_z$
is conserved. Here $r$ and $\theta $ refer to the usual polar
coordinates. Closed orbits occur if the radial and angular frequencies are
commensurate (see, for example \cite{LL}). Rewriting Eq.(\ref{ham}) as
\begin{equation}
H=(p_r ^2 +p_{\theta }^2/r ^2)/(2m)+m\omega ^2 r ^2/2-
\lambda \omega p_{\theta }^2/\hbar
\end{equation}
we obtain closed orbits if
\begin{equation}
t/s = 1/2 -\lambda p_{\theta }/\hbar
\end{equation}
with integers $t,s$. For instance, when $t/s=1/3,1/4,\ldots $ the
trajectory forms essentially a triangle, a square {\it etc.}
in the $\varrho-z-$plane; for $t/s=2/5$ we get the five star, and
so forth. The precise shapes of these geometrical figures depend
on the magnitude of $\lambda $ in that small values of
$\lambda $ produce polygons with rounded corners while larger
values yield loops at the corners as is illustrated in Fig.(1).
The appropriate scaling of $\lambda $ is given
by the kinematical constraint between the energy and the angular
momentum which reads
\begin{equation}
E+\lambda \omega p_{\theta }^2/\hbar \ge\omega |p_{\theta}|
\end{equation}
Here we stress again that for positive values of $\lambda $,
as used in actual applications, the value of the angular
momentum $p_{\theta }$ is not limited in its absolute value for given
energy; in particular, if $4 \lambda |E| >\hbar \omega $, arbitrarily large
positive or negative values may be assumed by $p_{\theta }$.

For the deformed system,
the orbits are Lissajous figures when $\lambda =0$. Closed orbits are
obtained for rational values of $b$. Our interest
is directed towards short periodic orbits.
Fig.(2a) displays the phase space structure
at $\varrho =0$ where the surfaces of section are taken. Note that
the whole region to the left and right, i.e.$\;$outside the
lines $z =\pm \sqrt{\hbar /(2\lambda m\omega )}$, is accessible
as long as the two lines do not intersect with the ellipse which
forms the other part of accessible phase space. If $\lambda $ is sufficiently
large to allow intersection of the lines with
the ellipse, the phase space becomes connected. In this case the part of
the ellipse which is to the left and right of the lines is
inaccessible. In Fig.(2b) surfaces of sections for $\lambda =0.03$
are given for four different orbits. With the choice of energy
$E=10$ and the frequency $\omega =1.5$ the two lines are outside
the ellipse. (For convenience we use units for which  $m=1$ and
$\hbar \omega =1$). The same pattern is 
obtained if $E$ and $\lambda $ are rescaled such that
$E \lambda =$const (see discussion below). The diagram of Fig.(2b) refers
to $b=2$, but the pattern as described prevails for $1<b\le 2$. Note that,
in accordance with the estimates below, the onset of chaos is discernible
just at the periphery of the ellipse, which is the region where $E_{\varrho }$
(the energy associated with the motion in the $\varrho $-variable) is
small. Typical trajectories
of periodic orbits are displayed in Figs.(3); Fig.(3a) corresponds to
the black dots on the far right and left of the phase diagram,
Fig.(3b) is the orbit sitting in the centre of the four islands,
and Figs.(3c) and (3d) are represented by crosses in Fig.(2b).
It was observed in \cite{hena} that there is a remarkable similarity between
short periodic orbits discussed above and the ones occurring in a WS potential
with increasing quadrupole deformation. In fact, the short periodic
orbits are of the same geometrical shape and occur in similar regions of the
surfaces of section. This similarity could indicate that shell structure
effects may be similar for the two potentials. A more careful analysis
taking into account stability and degeneracies of corresponding orbits
\cite{BM75,S76} goes beyond the scope of this paper.

For larger values of $\lambda $ hard chaos takes over quickly within
the ellipse, in particular when the two lines enter
the ellipse. We make two important observations: (1) regular
motion prevails far outside the ellipse which is the whole area to the
left of the left line and to the right of the right line.
These orbits may attain very large values of $\varrho $ and $z$;
also the variation of the angular momentum $p_{\theta }$ is unlimited
in principle. (2) The variation of $p_{\theta }$ ranges typically between
$-50$ and $+50$ for generic orbits inside the ellipse for our choice
of parameters. In Figs.(4) we illustrate typical surfaces of section and
trajectories for the two cases. The orbit on the bottom left of
Fig.(4) is representative for a regular orbit and its surface of section
produces the kind of tori lying completely outside the ellipse as
illustrated on the top left of Fig.(4); the irregular trajectory
on the bottom right of Fig.(4) produces the irregular dots inside and
outside the ellipse as shown on the top right; note the proximity to the
periphery of the ellipse of the dots outside the ellipse.

The characteristics of these two types of orbits are significant
when considering a
corresponding quantum mechanical calculation. The nearly regular motion for
very large values of $p_{\theta }$ is expected to lead to little mixing of
the class of states which is characterized by very large ($l>60$) and fairly
sharp values of the angular momentum. Also the quantum shell number is
still well defined for these states. These are the states which make
the quantum mechanical problem unbounded from below. Owing to their
negligible mixing to the states discussed in the following they
are virtually orthogonal upon the states in the chaotic region. Their
classically regular nature is reflected by the {\it de facto} existence of
good quantum numbers. The occurrence of chaos
is therefore not the result of unbounded motion, neither classically nor
quantum mechanically. However, the other class of states
is expected to have not only substantial expectation values
of the operator $(\vec L)^2$ (corresponding to $l\simeq 30$)
but in particular a variance of similar magnitude. Consequently,
shell structure is expected to be quite weak and
the concept of shell quantum number becomes ill defined. There is of
course a gradual transition between the two classes. The values given for
$l$ refer to $\lambda \le 0.02$. We note that these physically relevant
states obey the criterium $l\cdot \lambda <1$.
The results of the following section fully confirm such expectations.

We conclude the discussion of the classical motion by the results of a
perturbative treatment \cite{LiLi}
which was successfully employed in \cite{HNR}. When
$b>1$ we average over the angle $\theta _{\varrho }$ after rewriting the
Hamiltonian in terms of action and angle variables of the unperturbed problem.
This is sensible as long as the variation of $\theta _{\varrho }$ of the
exact motion is faster than that of $\theta _z$, which is ensured if
$E_{\varrho }\gg E_z$, where $E_{\varrho }$ and $E_z$ are the respective
energies residing in the $\varrho -$ and $z -$ motion. In principle, these
energies are not constants of motion, but the two degrees of freedom become
uncoupled within the approximation. In fact, we obtain after averaging
of Eq.(\ref{ham}) the effective Hamiltonian $H_{{\rm eff}}=
H_{\varrho }+ H_z$ with
\begin{eqnarray}
H_{\varrho }&=&{p^2_{\varrho }\over 2m}+{m\omega ^2\over 2}\varrho ^2\nonumber\\
   \\  \label{hr}
H_z&=&\alpha ({p^2_z\over 2m} + {m \tilde{\omega}_z^2\over 2})\nonumber
\end{eqnarray}
where
\begin{eqnarray}
\alpha &=& 1-2\lambda \frac{E_\varrho }{\hbar \omega}  \nonumber  \\
  \label{hz}  \\
  \tilde{\omega}_z &=&  \omega
\sqrt{({1\over b^2}-2\lambda \frac{E_\varrho }{\hbar \omega})/{\alpha}}
\nonumber
\end{eqnarray}
To ensure a real effective frequency $\tilde{\omega}_z$ in Eq.(\ref{hz})  we
read off an inequality which must be obeyed to render the approximation
valid. In view of $E\approx E_{\varrho }$ it may be written as
\begin{equation}
\label{bord}
\lambda E<{{\hbar \omega}\over 2b^2}.
\end{equation}
We note that the exact solutions are perfectly well represented by the motion
governed by $H_{{\rm eff}}$ if the inequalities are observed. We return
to the important inequality (\ref{bord}) in the next section.

So far we have concentrated our discussion upon $p_{\varphi }=0$ which
suffices for low values of $p_{\varphi }$. For larger values the pattern
changes, however, in that chaotic motion is suppressed to an increasing
extent for increasing values of $p_{\varphi }$. Stable periodic orbits
occur for large values of $p_{\varphi }$ at $\lambda $-values for which
the situation is fully chaotic at the lower $p_{\varphi }$-range.
An example is illustrated in Fig.(5). These orbits will of course prevail in
the corresponding quantum mechanical spectrum, when the full spectrum is
considered. We discuss in the next section that such levels
can lead to an overestimation of shell effects for the total energy of
large clusters.

\section{Quantum mechanical treatment}

As it was mentioned above for deformed system the obvious basis
uses the asymptotic
quantum numbers \cite{N,NM} which diagonalize the Hamiltonian
for $\lambda=0$. The boson operators are $a_z$ and $a_z^{\dagger }$
for the $z$-motion while for the two dimensional motion
perpendicular to the $z$-direction we use
\begin{eqnarray}
A_{\pm }&=&{1\over \sqrt{2}}(a_x\mp ia_y) \\
A_{\pm }^{\dagger }&=&{1\over \sqrt{2}}(a_x^{\dagger }\pm
ia_y^{\dagger }).
\end{eqnarray}
The conserved $z$-component of the angular momentum reads in this basis
\begin{equation}
L_z=n_+-n_-
\end{equation}
where $n_{\pm }$ are the occupation numbers referring to the
operators $(A_{\pm })^{\dagger }A_{\pm }$.
For a fixed eigenvalue of $L_z$, denoted by $\Lambda $,
we effectively have two quantum numbers defining the basis, i.e.
$n_z$ and $n_+$. This reflects the two degrees of freedom of the
corresponding classical situation.

The unperturbed ($\lambda =0$) spectrum is given by
\begin{equation}
E_{n_z,n_+}=\hbar \omega (2n_++1-\Lambda +{2n_z+1\over 2b}) \label{en}
\end{equation}
and the square of the angular momentum operator reads
\begin{eqnarray}
\label{vecl}
(\vec L)^2&=&{1\over 2}({1\over b}+b)\bigl( (2n_z+1)(2n_++1-\Lambda )
       -(a_z^2+(a_z^{\dagger })^2)(D+D^{\dagger })\bigr) \nonumber \\
       &+&{1\over 2}({1\over b}-b)\bigl( (a_z^2+(a_z^{\dagger })^2)
       (2n_++1-\Lambda )-(D+D^{\dagger })(2n_z+1)\bigr)\nonumber \\
       &+&(a_z^2-(a_z^{\dagger })^2)(D-D^{\dagger })-1+\Lambda ^2
\end{eqnarray}
where we used the combination $D=A_+A_-$.

It is now straightforward to set up the matrix of the full
Hamiltonian. Parity conservation is reflected by the fact that
the problem reduces into even or odd occupation numbers $n_z$.
The occupation numbers $n_+$ start with the positive value of $\Lambda $.
We stress that the general form for $(\vec L)^2$ as given by Eq.(\ref{vecl})
mixes states with $\Delta N = 0, 2, 4$ for $b\ne 1$.
As a consequence, the calculation cannot be restricted to a particular
shell number $N$ for the deformed case as it is usually done
in nuclear physics \cite{BM75,NR}. In
particular, the wave functions will be superpositions over a range of shell
numbers where the degree of the mixing increases with increasing deformation
and/or energy.

From Eq.(\ref{en}) we conclude that, if one decides
to truncate at $n_{+}^0=N_+$, the range of $n_z$ must extent up to
$N_z=[2b\cdot N_+]$ where $[...]$ denotes the integer part of a given number.
With this choice it is ensured that the unperturbed
basis is consistently represented up to
$E_{{\rm cons}}=(2N_++1-\Lambda +1/(2b))$ (in units of $\hbar \omega = 1 $),
in other words, all possible degeneracies are taken into account
up to $E_{{\rm cons}}$; for $b=1$ these degeneracies are just the possible
angular momentum values. Note that the range between $E_{{\rm cons}}$ and
$E_{{\rm max}}=(2N_++1-\Lambda +(2N_z+1)/(2b))$ cannot be omitted for
reasons of consistency, since $n_+$ and $n_z$ must run
{\it independently} to their respective maximal values.
These considerations are crucial in the present context as the mixing of
low lying levels with high angular momenta is essential.
It is instructive to discuss first the mixing of angular momenta
for $b=2$ and $\lambda =0$ where the eigenstates are exactly available.
In Table (1) we give the average
$\overline{L^2}=\langle n|(\vec L)^2|n\rangle $ and the variance
$\Delta L^2=\sqrt{\langle n|(\vec L)^4|n\rangle-(\overline{L^2})^2}$
for some selected positive parity levels for
$\lambda =0$ (recall $\Lambda =0$ and $\hbar \omega =1$).
Note in particular the large values of the variances which
often exceed those of the averages. The switching on of the $(\vec L)^2$
term not only lifts the degeneracies but brings about a strong mixing
among all levels. This becomes obvious in Table (2) where we display
corresponding quantities for $\lambda =0.01$. The appreciable
mixing of high angular momenta signals interaction of many
levels on a large scale which is also confirmed by the eigenvectors.
For the numbers chosen they pick up typically twenty and more
components larger than 0.1 from rather distant unperturbed states,
the eigenstates become strongly delocalized.
This mechanism requires a considerable extension of the basis
in mean field calculations even for relatively small strongly deformed
clusters.

The strong mixing between levels, irrespective of the basis chosen,
is related to the onset of chaos \cite{Gu90}, which is confirmed by the
nearest neighbor analysis of the spectrum. A particular sample is
illustrated in Fig.(6); a sample average would yield an almost perfect
match. This strong result
cannot be affected by a volume conservation condition
$\,\omega ^2\omega_z=const.$, which is used for the calculation of
Potential Energy Surfaces in order to find global and local
minima, which yield equilibrium shapes and shape isomers, respectively
(see e.g. \cite{Mon95,Ya95}).
In fact, while the effect of this constraint may change
the level ordering once the $(\vec L)^2$ term is included, it is observed
(e.g.$\,$\cite{NR}) that the ellipsoidal shape at large distortion becomes
unfavorable for any combination of filled single particle orbitals. In
addition, the increased level density at higher energy enhances the
interaction of the single particle states. Therefore, there is little
scope to find pronounced shell effects for a system with particle numbers
that fill shells beyond $N=4$ in the unperturbed basis and $b\ge 2$.
Likewise, for $N>20$, even a small deformation such as $b\approx 1.1$
does not support shell structure.

The levels listed in Table (2) are selected
according to the strength of their eigenvectors being essentially
concentrated in the upper fifth of their total length. We have chosen the
truncation of $n_+$ at $N_+=22$ which gives a $1035\cdot 1035$ total matrix
size. The states listed have half of their total norm exhausted within the
first 207 components. The spectrum of the states so selected is displayed
in Fig.(7a) as a function of
$\lambda $. In the top right corner we discern a distinctly irregular
pattern which is caused by the arbitrary criterion applied to select
the states. In fact, in this region the eigenvectors depend very
sensitively on the strength $\lambda $, since this is the region where level
repulsions occur on a large scale. Therefore, states from high lying
unperturbed levels will, owing to their coupling to lower
lying states, pick up considerable strength (0.5 for our choice) in their
first 207 components for a certain window of $\lambda $-values while
other states will undergo a similar change for a different window of
$\lambda $-values. In Fig.(7c) we display a section of the same states
without connecting the data points. Here it is clearly seen
that a few states maintain their identity for increasing
$\lambda $ while the majority of states is submerged in the morass of
a chaotic spectrum. To make this point even clearer, a stricter criterion
is used in Fig.(7d) where data are given for state vectors which
exhaust 80\% of their norm in the first
207 components. Obviously, only a few levels survive, in fact about
just the ones which mix weakly. Notice the fairly sharp
dividing curve between regular and irregular behavior in Fig.(7a)
which, according to the previous section, should be given
by the scaling law $E\lambda =$const.

When no selection is made there are many more levels with comparable
and even lower energies for $\lambda \ge 0.005$ as is illustrated in
Fig.(7b) where the complete spectrum is displayed. These additional states
mix to a lesser extent with the ones discussed above. The additional levels
come from higher lying unperturbed levels as seen in Fig.(7b) and
have even higher angular momentum components. These states are nearly
orthogonal to the states belonging to the chaotic spectrum; they correspond
to the states discussed in the previous section and form the space giving
rise to the negative part of the energy spectrum.
As a function of $\lambda $
the energies come down owing to the negative $(\vec L)^2$-term; accordingly,
the corresponding eigenvectors have their bulk components at higher labels.
They are the quantum mechanical counterpart of the type of classical
orbits displayed on the left column of Fig.(4). Traditionally these states
are pushed up in their energy by the {\it ad hoc} adding of the mean value of
$L^2$ which is, for small deformation, a simple expression in the shell
number \cite{Cl85,Ya95}. This remedy does not affect the eigenvectors, it
would change the spectrum but not its statistical properties.
The basic reason for this to hold is the fact that the states whose energies
tend to minus infinity do not couple with the physically relevant
states as the irrelevant states have very large angular momentum while the
relevant states have a mixture of only moderate angular momenta.
Since the concept of shell numbers breaks down, the mean value of $L^2$ can
no longer be expressed in terms of a shell number and adding it to the
spectrum becomes doubtful. On the basis of our findings we rather ignore
the irrelevant states.

For completeness we report corresponding results for $b=1.1$.
In Table (3) we present results for $\lambda =0.1$. To demonstrate
the only slightly broken symmetry for the near spherical case
this value is chosen much larger than the one usually applied in either
nuclear physics or metallic clusters. As a result, irrelevant states with
large mean values of angular momentum
($l > l_{{\rm crit}}=1/\lambda$) have negative energies.
All states are fairly good eigenstates of angular momentum with
much smaller spreading $\Delta L^2 $ as is expected for this mild
deviation from unity of $b$. In most cases the expectation value of
$(\vec L)^2$ is very close to an eigenvalue $l(l+1)$, we have
indicated the corresponding $l$-values in brackets.
The eigenvectors are fairly well localized, there is only
limited mixing. A near Poisson distribution is
found for the nearest neighbor distribution indicating the
near-integrability of the $b=1.1$ case. Note, however, that these statements
are limited to about the first three hundred levels.

Based upon the inequality $E_z\ll E_{\varrho}$ we obtain from Eq.(\ref{bord})
the border line $E\lambda \simeq 1/(2b^2)$ between regular motion and expected
chaos. A similar dividing line should be expected for the quantum
mechanical problem. It appears that the relation $E\lambda \simeq 1/(4b^2)$
can serve as an empirical dividing line between order and chaos as long
as $b>1$. The difference between Eq.(\ref{bord}) and the empirical
quantum mechanical result is due to the approximation; also, such border
lines are not expected to coincide exactly for the classical and quantum
mechanical case. We may combine this relation with
$E\approx l(1-\lambda l)\, (n_r=0, \hbar \omega=1,\,N$ and $l$
sufficiently large) to obtain the estimate
$\lambda l \simeq 1/2$ as a border line at around $b=1$. Therefore, values of
$l$ and $\lambda$ which obey the condition $1/2\leq \lambda\cdot l \leq 1$
correspond to chaotic motion. (Note that, since outside the chaotic region
the values of the angular momentum are almost sharp, values of $l$ can
be used to delineate different regions even though it is a dynamical
variable.) As a consequence,
the values used for $\lambda$ in nuclear physics, when
applied for large metallic clusters, lead to a strong mixing of
levels at the high end of the spectrum (see Fig.(7a,7b)) even for values
of $b$ close to unity. Hence, shell structure is expected to be weakly
pronounced for a large particle number even for mild deformation.
For $b=2$ the emerging shell structure is confined to the very low
end of the spectrum and to rather small values of $\lambda $.

The results found for the specific examples $b=1.1$ and $b=2$
do not change for negative parity states or for $\Lambda >0$.
The spectrum for larger values of $\Lambda $ nicely slots in with
the spectrum for $\Lambda=0$ to produce typical shell
structure as long as $b$ is close to unity. At the end of Section 3 we
pointed out that a regular pattern begins to emerge for
very large values of $p_{\varphi }$ even when $b=2$.
Note also that, for large values of $\Lambda $ (the quantum
mechanical value of $p_{\varphi }$), $(\vec L)^2$ becomes diagonally dominant
as seen from Eq.(\ref{vecl}) thus leading to lesser pronounced level
repulsion in line with the increasingly regular classical situation. While
levels of large $\Lambda $ are submerged in the morass
of a chaotic spectrum pertaining to lower $\Lambda $-values, they are
expected to yield discernible shell structure in a Strutinsky type
analysis. In calculations by \cite{RBH,RFB}, levels belonging to rather
high shells occur below the Fermi surface and have to be taken into account
in the Strutinsky averaging procedure; pure oscillator states $(\lambda=0)$
are used for the upper part of the spectrum. Even for relatively small
clusters with sizes up to $A=270$ atoms \cite{RFB} the convergence
criterium for the Strutinsky's shell correction method \cite{Str}
required inclusion of spherical shell numbers up to $N=16$. In essence,
the occurrence of large $\Lambda $ is associated with very high shell
numbers, a region where the applicability of the
model is no longer certain. For example,
in \cite{RBH} it was found that for large spherical clusters
$(A=676$ and $832)$ the energy minima are more pronounced than in the
KS calculations, which is an artifact. In other words, while in
the axial MHO potential the re-emergence of order and hence shell structure
is a fact for large $\Lambda $, its physical impact cannot be assessed
from this model alone. Moreover, even small non-axial deformations, which
are discernible for small clusters \cite{Ya95,RFB}, will destroy shell
effects for large clusters, since $L_z$ is no longer a constant of motion,
i.e. $\Lambda $ is no longer a good quantum number.

\section{Summary}

Using the classical analysis of single particle motion in the MHO
potential our attention is focused on the
distinction of regions in phase space of predominantly regular or chaotic
motion, when the deformation $b$ and the term $(\vec L)^2$ act
simultaneously.
Important similarities, not obvious at first glance,
with regard to the shortest periodic orbits which define the gross shell
structure of the single particle spectrum
in the MHO potential and in the Woods-Saxon potential
have been found previously \cite{hena} .
Typical trajectories of periodic orbits are displayed in Fig.(3).
Both cases give rise to chaotic motion when the deformation in the
Woods--Saxon and the $(\vec L)^2$-term in the MHO potential are
turned on. The chaotic behavior is of lesser significance,
as long as the interest is focused only on the lower end of the spectrum
and the deformation is fairly mild, that is
for small metallic clusters. For large
metallic clusters, however, with the much larger particle numbers, a higher
range of the spectrum becomes relevant; depending on the deformation,
chaotic behavior may well interfere with the search for shell structure
in this region.

Regular and chaotic behavior are essentially separated by the curve
$E\lambda \simeq \hbar \omega /(4b^2)$. When  $b>1$
and $\lambda =0$, the problem is regular and so is the variation of the
classical angular momentum; in fact the variation is determined by
the frequencies $\omega $ and $\omega /b$. If $\lambda $ is switched on,
the variation becomes quickly irregular, but more importantly, the range
of variation increases considerably with increasing $\lambda $ as it should be
expected, since the term $\lambda (\vec L)^2$ adds angular momentum.
In quantum mechanics, switching on the $(\vec L)^2$ term not only lifts
the degeneracies but brings about a strong mixing among all levels
of the unperturbed basis. Therefore, the unavoidable truncation of the
basis in such a situation must be handled with great circumspection to
avoid the occurrence of spurious shell structure.
While it is common wisdom that questions of truncation
are delicate, we stress that the classical finding, which is the large
fluctuation of the angular momenta, provide essential guidance to the
problem of the truncation.

From Figs.7 and from the relation $E\lambda \simeq 1/(4b^2)$  
($\hbar \omega =1$)
we conclude that shell structure is possible at higher energies
(larger number of particles) only for accordingly small values of $\lambda$.
For $\lambda \ge 0.007$ (which is smaller by one order of magnitude than
the typical ones) and  $b=2$ hard chaos occurs already for $E\geq 10$
which is seen from the Wigner distribution of the nearest neighbor statistic.
Shell structure is retrieved for lesser deformation at the same values of
energy and $\lambda $. One of the basic conclusions of our analysis is that
using the MHO model for metallic clusters with a coupling strength
$\lambda $ similar to that in nuclear physics would conflict, at
least for larger deformations $b$, with observation of
shell structure. Within the model considered we
may speculate that large particle systems tend to restore the near-spherical
symmetry to allow for shell structure which basically means increased
stability for the appropriate particle number. There seem to be indications
to this effect in calculations of the equilibrium shapes for sodium
clusters with $A\leq 300$ (see, for example \cite{RFB}) using the MHO.
We conclude that the MHO potential cannot yield
shell structure for $b\ge 2$ and shell number larger than about 4, or
$b\simeq 1.1$ and shell number larger than about 20. Energy minimization is
therefore expected to yield decreasing deformation for increasing particle
number.

When the model was introduced originally in nuclear physics
\cite{N}, neither the unbounded motion nor the inherent chaotic nature
was of interest. The $(\vec L)^2$ term has been introduced to modify the
nuclear potential with an effective A-dependence. Its chaotic nature
was masked at the time, since only $A\leq 200$ was of interest.
The transfer of a well established nuclear structure model to the new
physical situation of metallic clusters requires a re--assessment of the
phenomenological model, since it entails systems of a considerably larger
number of particles, where even a moderate deformation can have a pronounced
effect. The connection
between shell structure phenomena in the quantum mechanical spectrum and
ordered motion in the analogous classical case leads to the
conclusion that the model is useful but only in a restricted domain.
Our analysis shows that the two aspects, the unboundedness and chaotic
motion, are independent of
each other. In fact, the regular classical motion for large values of
$(\vec L)^2$  ($l>l_{{\rm crit}}=1/\lambda$)
gives rise to quantum mechanical states which have virtually
sharp values of $l$ and are essentially orthogonal upon the chaotic states
which, in turn, are characterized by the inequality
$1/2 \leq l\lambda \leq 1$. The other regular part of the spectrum which is
displayed in Fig.7 is obtained for $l\lambda <1/2$.

The inclusion of higher multipoles into the potential for large system
cannot change this finding. The chaotic nature inferred by adding
the term $(\vec L)^2$ destroys the shell structure which is
produced in a subtle way by the octupole deformation when added to the
quadrupole deformed HO \cite{H94}. The hexadecapole deformation, when
added to the quadrupole deformed HO, induces shell structure with a pattern
similar to that of a lesser deformed quadrupole HO \cite{H95}. And the
combination yields shell structure for specific values of the octupole and
hexadecapole strength \cite{HNR}. Since the $(\vec L)^2$ term
dissolves all bunching of levels in a deformed potential at higher energies,
i.e.$\;$for large particle numbers, shell structure can be restored only
for a near spherical shape. This property is
shared with the effect of the hexadecapole term mentioned above. In fact,
similarities between the two terms have been pointed out in \cite{BM75}.
Other shapes of potentials \cite{FP} seem to indicate that adding higher
multipoles gives rise to shell structure for large particle numbers. It is
a challenge to understand such claims in terms of periodic orbits. Work
towards this aim is in progress.

\vspace{0.5in}
This project has been supported by the Foundation for Research
Development of South Africa.

\newpage
\begin{center}
\begin{tabular} {|c|c|c|c|}   \hline
n & $E_n$ & $\overline{L^2}$ & $\Delta L^2$\\ \hline
1 & 1.25 & 0.25 & 1.35 \\
2 & 2.25 & 5.25 & 5.73 \\
3 & 3.25 & 10.3 & 11.5 \\
4 & 3.25 & 2.75 & 4.85 \\
5 & 4.25 & 17.8 & 15.7 \\
6 & 4.25 & 15.3 & 17.3 \\
7 & 5.25 & 32.8 & 29.1 \\
8 & 5.25 & 20.3 & 23.1 \\
9 & 5.25 & 5.35 & 8.78 \\
17 & 8.25 & 80.3 & 70.7 \\
18 & 8.25 & 77.8 & 69.9 \\
19 & 8.25 & 42.7 & 38.2 \\
20 & 8.25 & 35.3 & 40.6 \\
26 & 10.25 & 130 & 115 \\
27 & 10.25 & 113 & 99 \\
28 & 10.25 & 108 & 97 \\
29 & 10.25 & 56 & 50 \\
30 & 10.25 & 45 & 52 \\
\vdots & \vdots & \vdots &\vdots \\ \hline
\end{tabular}
\end{center}
{\bf Table (1)}.
Average values and variances of the operator $(\vec L)^2$ for a few low
energy states for $b=2$ and $\lambda =0$.
\newpage
\begin{center}
\begin{tabular} {|c|c|c|c|}   \hline
n & $E_n$ & $\overline{L^2}$ & $\Delta L^2$\\ \hline
1 & 1.25 & 0.29 & 1.47 \\
2 & 2.19 & 5.85 & 6.51 \\
3 & 3.12 & 13.3 & 14.7 \\
4 & 3.22 & 2.82 & 5.24 \\
5 & 4.04 & 24.0 & 28.9 \\
6 & 4.06 & 18.7 & 16.3 \\
7 & 4.85 & 43.8 & 47.3 \\
8 & 4.96 & 37.0 & 39.7 \\
9 & 5.19 & 6.41 & 9.80 \\
13 & 5.91 & 32.3 & 36.6 \\
14 & 6.25 & 1657 & 1862 \\
15 & 6.30 & 2163 & 1973 \\
16 & 6.50 & 1947 & 1963 \\
27 & 7.76 & 51.3 & 117 \\
28 & 7.93 & 1504 & 1740 \\
29 & 7.95 & 1413 & 1810 \\
30 & 8.13 & 836 & 1479 \\
\vdots & \vdots & \vdots &\vdots \\ \hline
\end{tabular}
\end{center}
{\bf Table (2)}.
Same as Table (1) but for $\lambda =0.01$.

\newpage
\begin{center}
\begin{tabular} {|c|c|l r|c|}   \hline
n & $E_n$ & $\overline{L^2}$&$(l)$ & $\Delta L^2$\\ \hline
1 & -28.1 & 506 &(22) & 9.6 \\
2 & -21.5 & 420 &(20) & 9.0 \\
3 & -19.5 & 420 &(20) & 9.9 \\
4 & -15.6 & 342 &(18) & 8.4 \\
9 & -6.7 & 272 &(16) & 9.7 \\
10 & -6.2 & 210 &(14) & 7.4 \\
17 & -0.17 & 110 &(10) & 6.7 \\
18 & 1.1 & 156 &(12) & 8.7 \\
20 & 1.45 & 0.01 &(0) & 0.2 \\
21 & 1.87 & 71.6 &(8) & 6.9 \\
22 & 1.89 & 110 &(10) & 7.6 \\
23 & 2.7 & 6.7 &(2) & 4.4 \\
24 & 2.9 & 39 &(6) & 10.5 \\
25 & 3.0 & 156 &(12) & 9.7 \\
26 & 3.3 & 22.8 &(4) & 8.5 \\
27 & 3.4 & 0.2 &(0) & 1.7 \\
\vdots & \vdots & \vdots &\vdots &\vdots \\ \hline
\end{tabular}
\end{center}
{\bf Table (3)}.
Same as Table (1) but for $b=1.1$ and $\lambda =0.1$. Note that for this
small deformation the levels are nearly good eigenstates of $(\vec L)^2$
with averages close to $l(l+1)$ and small variances.
\newpage
\vspace{1cm}
\centerline{\bf Figure Captions}

{\bf Fig.1} Typical simple periodic orbits for $b=1$.
The orbits with the loops at the corners are for $\lambda
{\; \raisebox{-0.4ex}{\tiny$\stackrel{{\textstyle>}}{\sim}$}\;}
\hbar \omega /(4E)$
while the other two are for $\lambda  = \hbar \omega /(25E)$.

{\bf Fig.2} Surfaces of section in the plane $\varrho =0$. In (a) the
dependence on $\lambda $ of accessible phase space (shaded) is illustrated;
left: $\lambda <\hbar \omega/(4E)$, right: $\lambda >\hbar \omega/(4E)$.
In (b) four orbits of different initial conditions are displayed
for $\lambda <\hbar \omega/(4E)$.

{\bf Fig.3} Typical short periodic orbits in the $\varrho -z-$plane.
Reference to Fig.2 is made in the main text. Figs.(c) and (d) are
self-tracing orbits.

{\bf Fig.4} Two orbits for the same choice of parameters
($b=2,\lambda =0.01$). The left hand column displays the surface of section
and the corresponding orbit for a regular orbit while the right hand column
gives the respective plots for a typical chaotic orbit.

{\bf Fig.5} Two orbits for the same choice of parameters and initial
conditions but for different values of $p_{\varphi }$.
The left hand column displays the surface of section at $\varrho =8$
and the corresponding orbit for a regular orbit at $p_{\varphi }=61$
while the right hand column gives the respective plots for a typical chaotic
orbit at $p_{\varphi }=1$.

{\bf Fig.6} A nearest neighbour distribution of the unfolded spectrum for
$b=2$ and $\lambda =0.01$.

{\bf Fig.7} (a) Lower end (about 6\% of all levels) of the energy spectrum
in units of $\hbar \omega $ as a function of the strength $\lambda $ for
states whose eigenvectors exhaust 50\% of the norm in the first fifth of the
total length. The dividing line $E\lambda =1/(4b^2)$ can be nicely discerned.
(b) Lower end of the spectrum for all levels. (c) A section of (a).
(d) Same as (c) but only for states whose eigenvectors exhaust 80\% of the
norm in the first fifth of the total length.

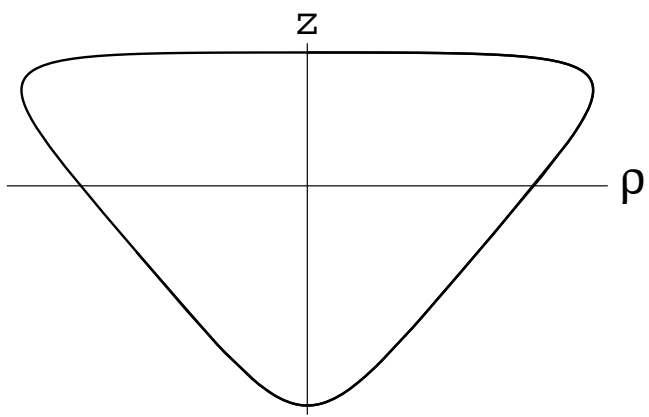
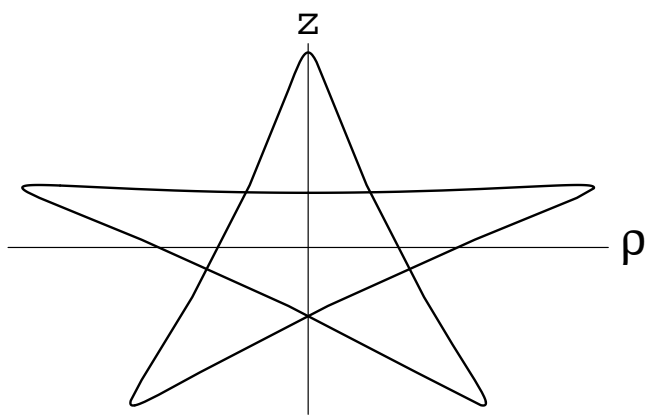
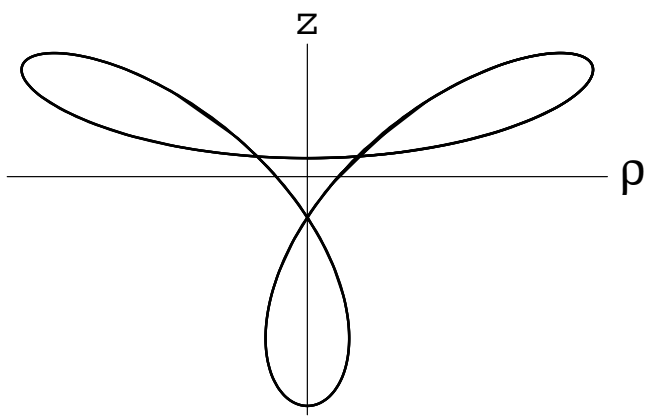
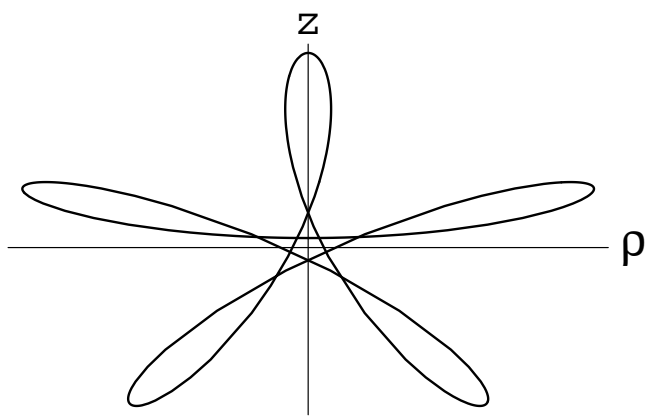
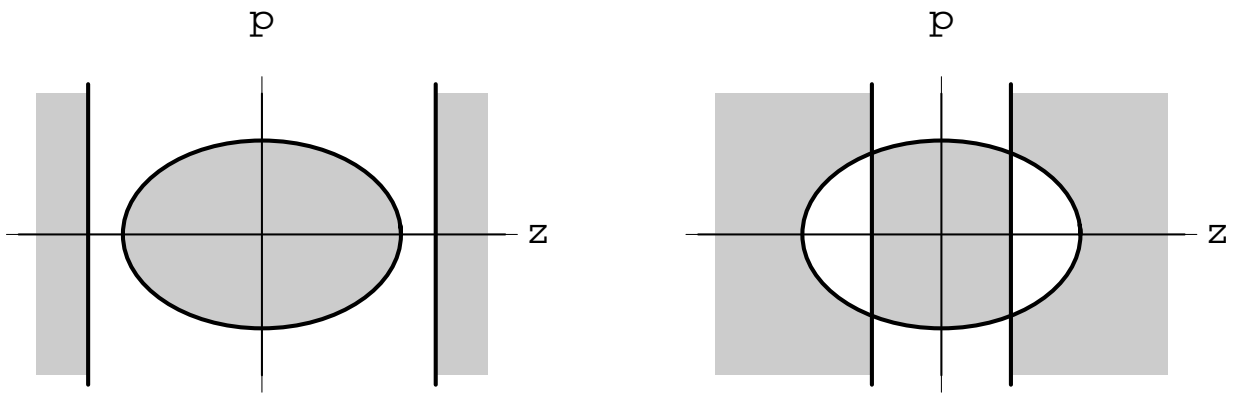
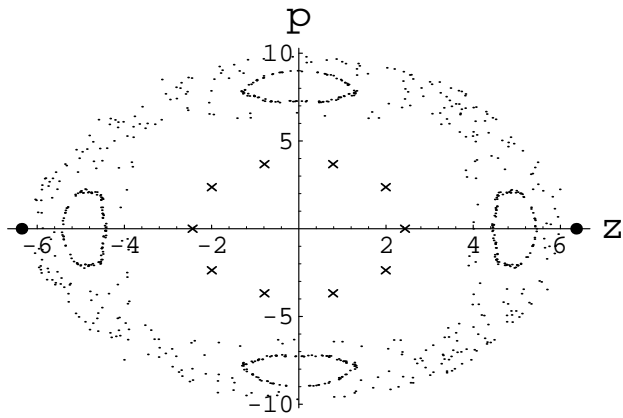
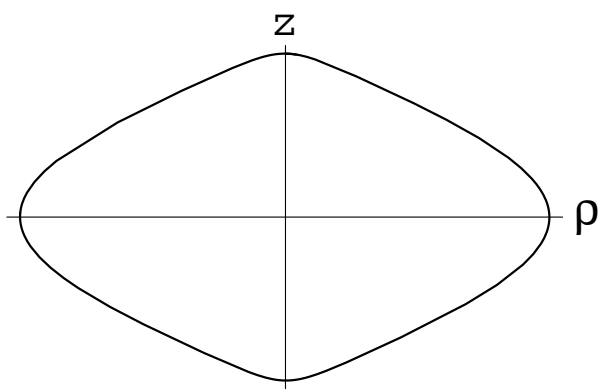
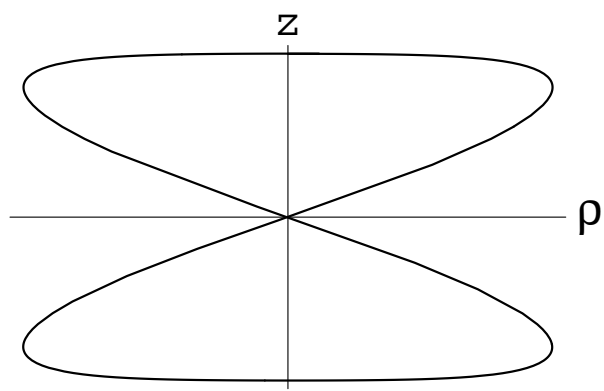
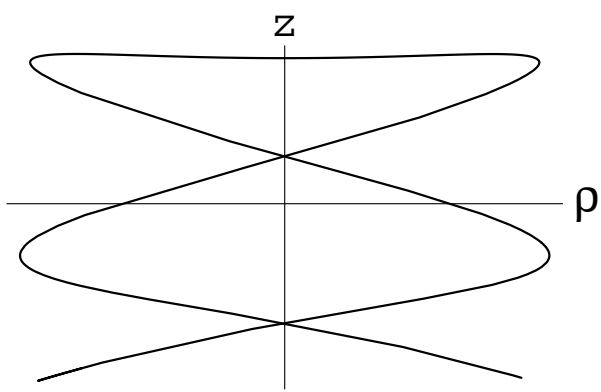
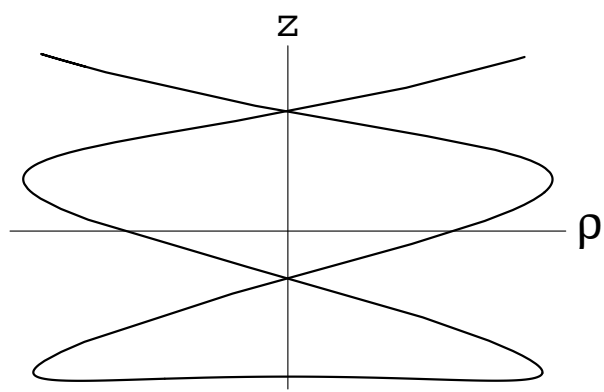
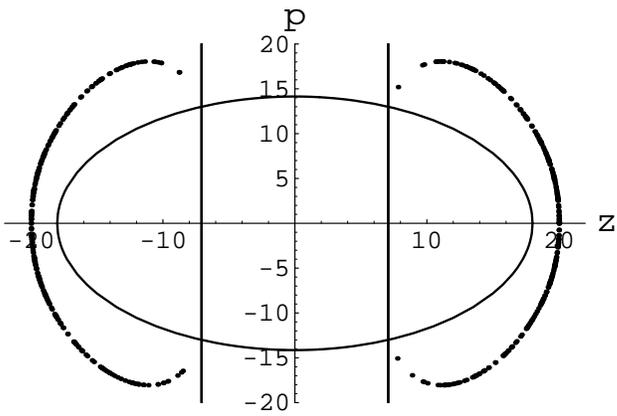
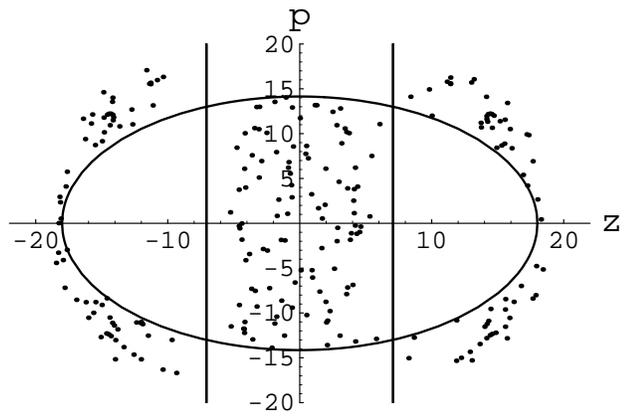
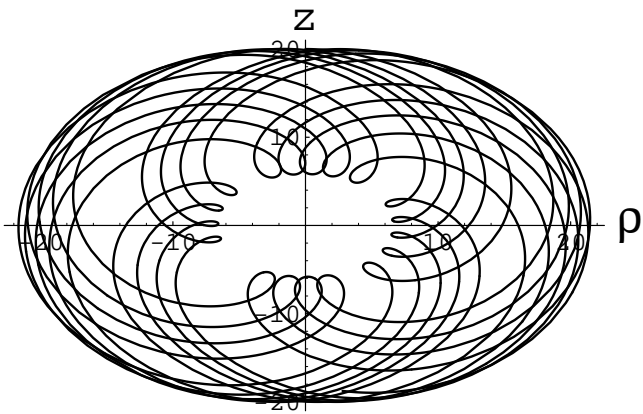
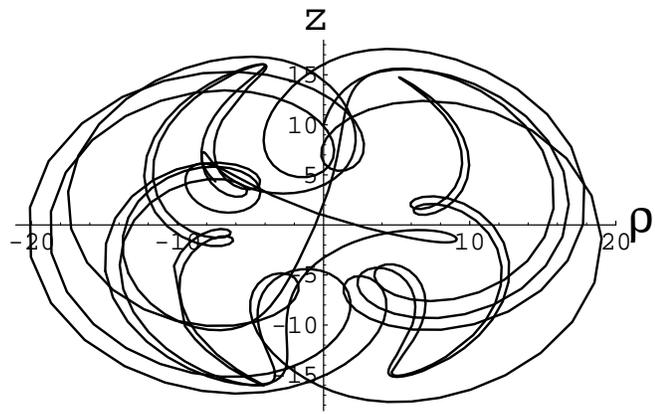
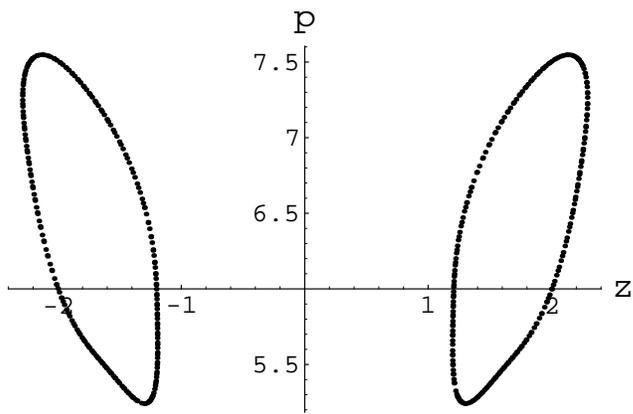
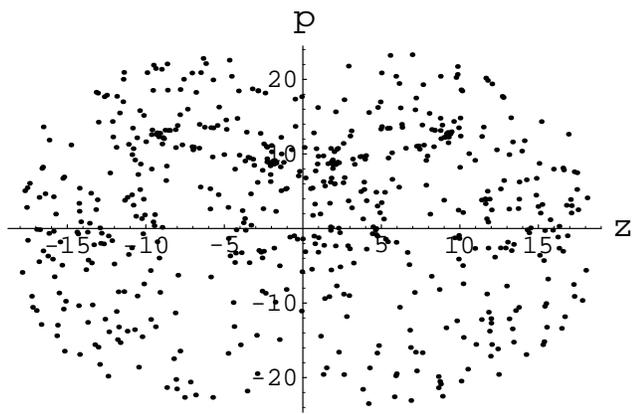
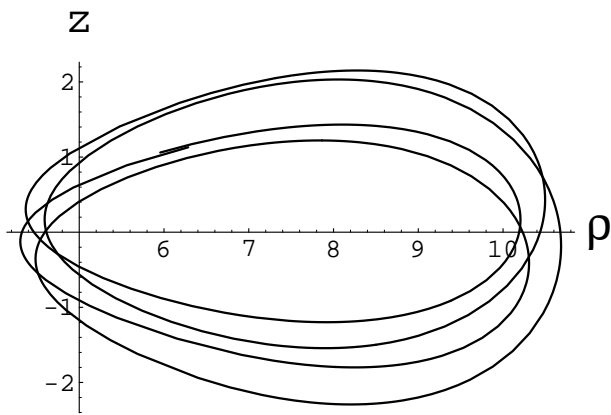
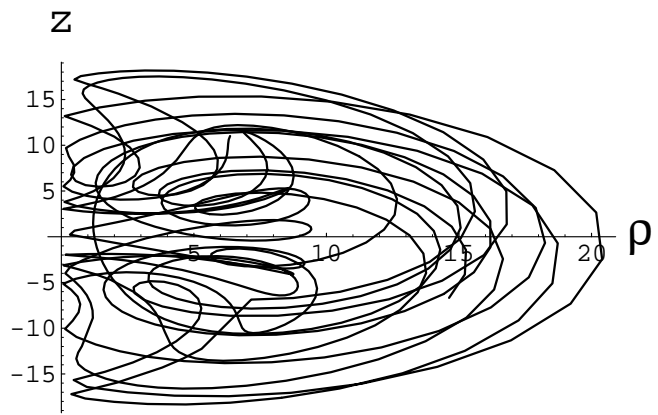
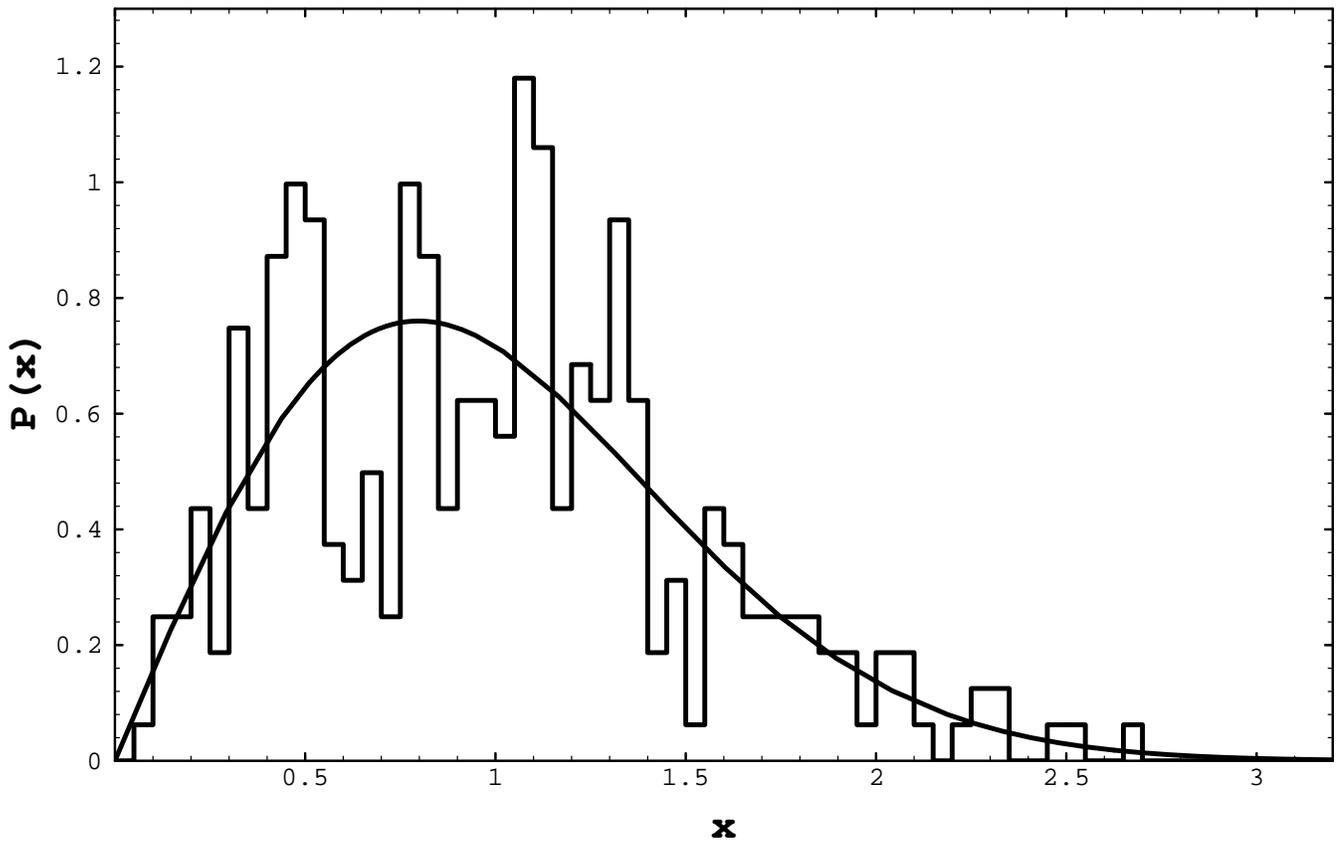
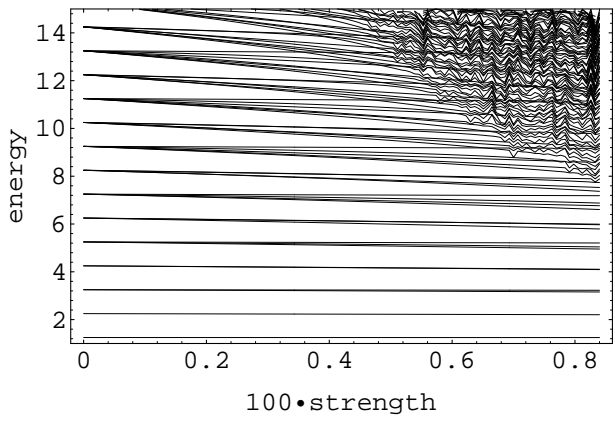
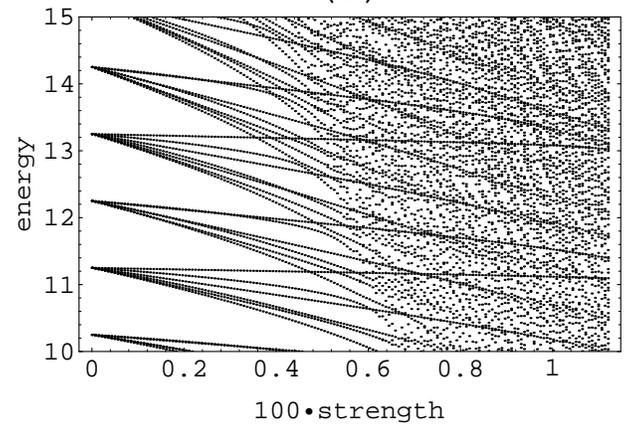
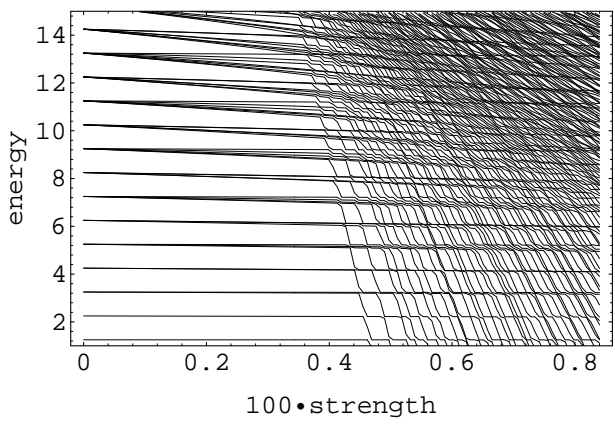
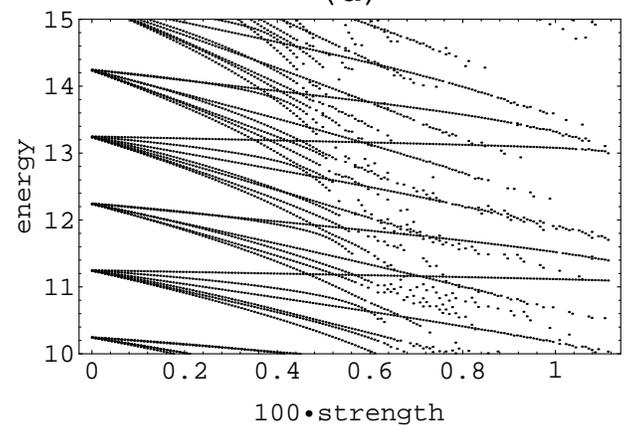
\end{document}